\documentclass{epl}
\usepackage{epsfig}

\title{Dynamic Modeling of the Electric Transportation Network}
\shorttitle{Dynamic Modeling of the Electric Transportation Network}
\author{Alessandro Scir\`{e}\thanks{E-mail: \email{scire@imedea.uib.es}},
Id\'an Tuval, V\'{\i}ctor M. Egu\'{\i}luz}
\institute{Instituto Mediterr\'aneo de Estudios Avanzados IMEDEA (CSIC-UIB),
E07122 Palma de Mallorca, Spain}
\shortauthor{A. Scir\`e \etal}


\begin{document}

\maketitle

\begin{abstract}
We introduce a model for the dynamic self-organization of the electric grid.
The model is characterized by a conserved magnitude, {\em energy}, that can
travel following the links of the network to satisfy nodes' {\em load}. The load
fluctuates in time causing local overloads that drive the dynamic evolution of
the network topology. Our model displays a transition from a fully connected
network to a configuration with a non-trivial topology and where global failures
are suppressed. The most efficient topology is characterized by an exponential
degree distribution, in agreement with the topology of the real electric grid.
The model intrinsically presents self-induced break-down events, which can be
thought as representative of real black-outs.
\end{abstract}

\section{Introduction} The electric grid is a critical infrastructure for our
economy and society. Recent events, ranging from the large-scale blackouts a
few years ago to the California crisis today \cite{california},  highlight the
need to enhance the insight on the electric grid, complementing the traditional
technological analysis \cite{IEEE} with new transversal points of view. Our
approach is to study the electric grid macroscopic behavior rather than to
dissect individual events. At that macroscopic scale, the electric grid
exhibits behaviors typical of complex systems. For instance, on the basis of
15-years time series of transmission system black-outs of the U.S. electric
network \cite{carreras,carreras04}, it has been proposed that the electric grid
may be a self-organized critical system, operating at or near a critical point.

A signature of the electric grid is that it can be represented as a complex
network, where nodes are the generators and the links the transmission lines.
Recent research in complex networks has shown that a detailed knowledge of the
topology of a communication or transportation network is essential for the
understanding of cascading failures
\cite{Strogatz01,Albert02,Dorogovtsev02,Newman03}. While some of these studies
have focused on the topological robustness of the underlying network to random
failures and targeted attacks \cite{Crucitti04}, other research have considered
dynamic processes on static networks \cite{PastorSatorras01,Eguiluz02}. However,
these studies do not consider the network as a dynamic entity whose evolution is
driven by the action of the nodes
\cite{Eguiluz00,Ito02,Marsili03,Anghel04,bianconi04,Zimmermann04}. In this
Paper, we present a dynamic model aiming to describe the growth and evolution of
a transportation network. The network growth relies on the need of resource
distribution in a heterogenous environment.

\begin{figure}[tb]
\centerline{\twoimages[width=0.4\columnwidth]{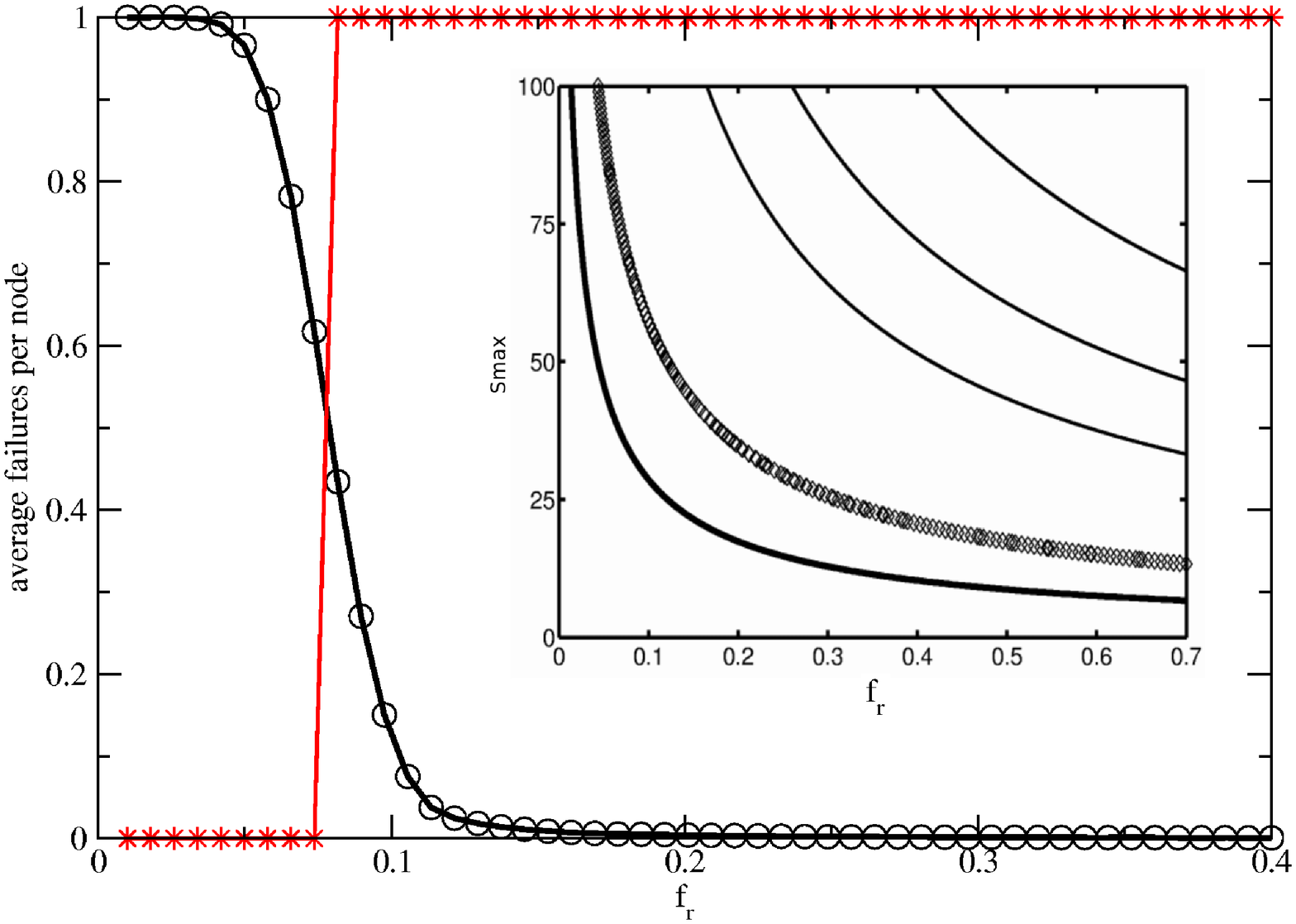}{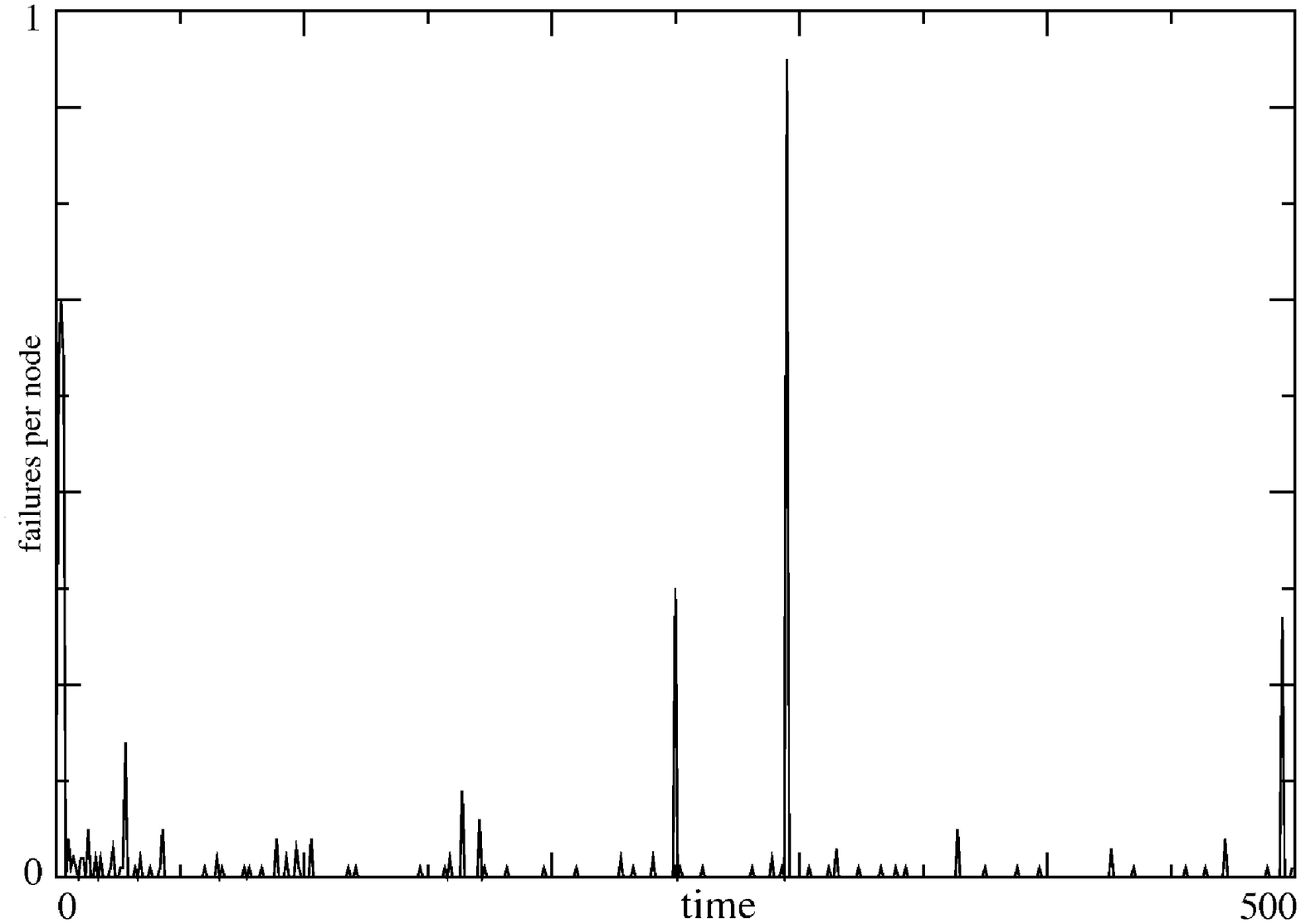}}
\caption{\label{transicion} (Left panel) The average failure
events per node (circles) and the integer part of
$\epsilon/\lambda$ (stars) versus $f_r$. $s_{\rm max}=100$,
$\mu=1/2$, $N_{step}=2500$, N=2500. Inset. Thick line:
$\varepsilon = \lambda$; diamonds: $\varepsilon_{\rm isolated}$;
solid lines: other iso-energy lines $\varepsilon=5,10,15$,
respectively. (Right panel) Temporal evolution of the total number
of failures per node. $f_r=0.3$.}
\end{figure}

\section{The Model} We consider $N$ dynamic elements located at the nodes of a
two-dimensional square lattice. Each element $i$ is characterized by its size
$s_{i}$, drawn from a probability distribution $p(s)$. To each element $i$ are associated
two dynamic variables: the {\em load} (energy consumption) $l_i^t$, and the
{\em supply} (available energy) $f_i^t$. We assume that (i) the load of element
$i$ is described by $l_i^t = m_i + \sqrt{m_i}\xi_i^t$, where $m_i$ is a constant value
and $\xi_i^t$ represents a fluctuation term; and (ii) initially at $t=0$ the
available energy and the constant load at each element are proportional to its
size $s_i$
\begin{eqnarray}
f_i^0 &=& f_r s_i   \label{fm1}     \\
m_i &=& m_r s_i~,
\label{fm2}
\end{eqnarray}
where $f_r$ and $m_r$ are constant values and in general $f_r > m_r$. Starting
from an initial condition where each element is isolated, the network grows as follows. At each time step:

\begin{enumerate}

\item If the load overcomes the supply at element $i$ ($f_i^t < l_i^t$), a {\em
failure} occurs.

After a failure, the failing element $i$ chooses a target neighbor through a
wiring strategy. Following the empirical results observed in communication
networks with spatial constraints \cite{Yook02}, the target node $j$ is chosen
such that it maximizes the function
\begin{equation}
\pi(i,j)=\frac{s^{\beta}_j}{d^{\gamma}_{ij}}~,
\label{preferential}
\end{equation}
where $d_{ij}$ represents the Euclidean distance between the two elements, and
the exponents $\beta$ and $\gamma$ indicate the preference for size and
proximity, respectively, in the wiring.

\item The supply at element $i$ is updated according to
$f_i$.
\begin{equation}
f^{t+1}_{i}=\left\{ \begin{array}{cc}
f^{t}_{i} & \mbox{if } k_i=0                        \\
l^{t}_{i}+\sum_{j\in{\cal V}(i)} \frac{f^{t}_{j}-l^{t}_{j}}{k_j} & \mbox{if } k_i \ne 0
\end{array}
\right.
\label{fmap}
\end{equation}
where $k_j$ is the number of links possessed by the element $j$, $j$ running
over the neighbors of the element $i$, ${\cal V}(i)$. The prescription given by
Eq.~(\ref{fmap}) allows the energy to be distributed through the links of the
network as it is needed, depending on the instantaneous load of each linked
element, making next overloads more unlikely.
\end{enumerate}

\begin{figure}[tb]
\onefigure[width=0.77\columnwidth]{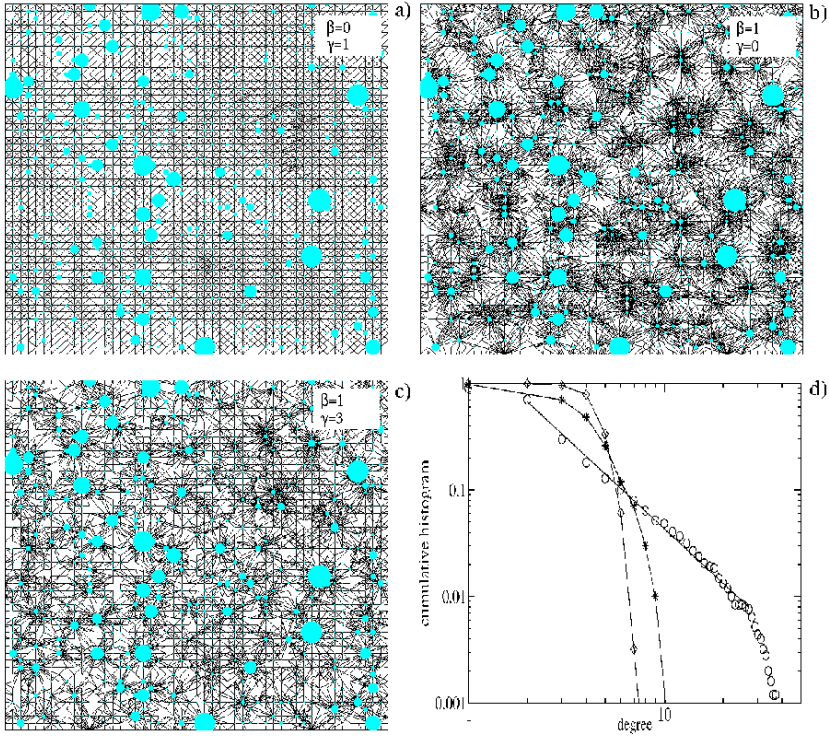}
\caption{\label{Topology}(a-c) Network structure; d) Degree distribution
corresponding to the network displayed in a) (diamonds), b) (circles), and c)
(stars). Lines correspond to fitted curves.}
\end{figure}

Each $N_{steps}$ time steps, the links are actually set and the network is
consequently updated. The choice of the time step for the network construction
is a further degree of freedom of our model. In real electric grids, the time
scales of the fluctuating demand are much faster than the time scales at which
the network is modified. Therefore, we choose to update the network each
$N_{steps}\gg 1$ time steps of the local dynamics.

The initial total supply $E^0$ and the total load $L^0$ in the system are
given by
\begin{eqnarray}
E^0 &=& \sum_i f_r s_i = N f_r \Lambda_{1}^{P}  \label{energy}  \\
L^0 &=& \sum_i m_r s_i + \sqrt{m_r s_i}\xi_i^0 = N m_r \Lambda_{1}^{P} + N\sqrt{m_r}\Lambda_{2}^{P}\langle \xi \rangle~,
\end{eqnarray}
where $\Lambda_{1}^{P}$ and $\Lambda_{2}^{P}$ are two coefficients that depend only on the geographic
distribution of city sizes $P(s)$. Under the transformation given by
Eq.~(\ref{fmap}), the total energy $E^0$ is conserved. Thus in the remainder we
will remove the superscript.

To model the actual geographic distribution of city sizes, the distribution of $s_i$ follows a Zipf law~\cite{zipf} of the form $P(s) = A s^{-2}$ in the interval $(1,s_{\rm max})$ and $A=s_{\rm max}/(s_{\rm max}-1)$ is the normalization coefficient. Fluctuations $\xi_i$ are described either by the logistic map in the chaotic regime, or by white noise \cite{ruido}.

\begin{figure}[tb]
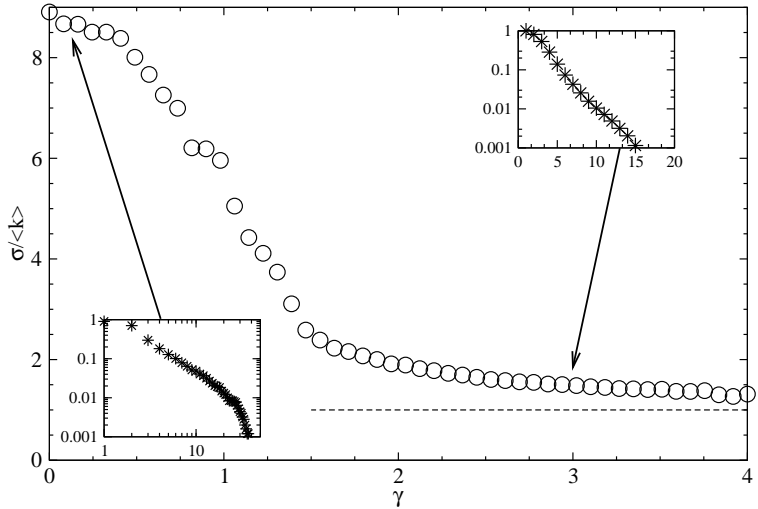

\centerline{\onefigure[width=0.7\columnwidth]{fig3N.eps}}
\caption{\label{sigma}
The normalized standard deviation $\sigma/\langle k \rangle$ of the degree distribution versus
$\gamma$ for $\beta=1$. Insets: degree distribution corresponding to $\gamma=0$
and $\gamma=3$, with fixed $\beta=1$.}
\end{figure}

The average energy $\varepsilon$ and average load $\lambda$ per node
in the system are now given by
\begin{eqnarray}
\varepsilon &=& \frac{E}{N} = f_r \frac{s_{\rm max}\ln{s_{\rm max}}}{s_{\rm max}-1}
                            \equiv f_r \Theta (s_{\rm max}) \\
\lambda &=& \frac{L}{N} = m_r \Theta (s_{\rm max}) + \frac{1}{4}\sqrt{m_r}\frac{\sqrt{s_{\rm max}}}{\sqrt{s_{\rm max}}+1}
            \equiv m_r \Theta (s_{\rm max}) + \frac{1}{2}\sqrt{m_r}\Phi(s_{\rm max}) \label{totalenergy}~.
\end{eqnarray}
It is worth noting that at any given time the maximum possible load in
the network is $\lambda_{\rm max} = m_r \Theta (s_{\rm max}) + \sqrt{m_r}\Phi(s_{\rm max})$. Thus for a
system composed by isolated nodes following the same dynamics, the total
energy that guarantees that all the nodes have access to the energy they
need is
\begin{equation}
\varepsilon_{\rm isolated} = \lambda_{\rm max} = m_r \Theta (s_{\rm max}) + \sqrt{m_r}\Phi(s_{\rm max})~.
\end{equation}

\begin{figure}[tb]
\onefigure[width=0.6\columnwidth]{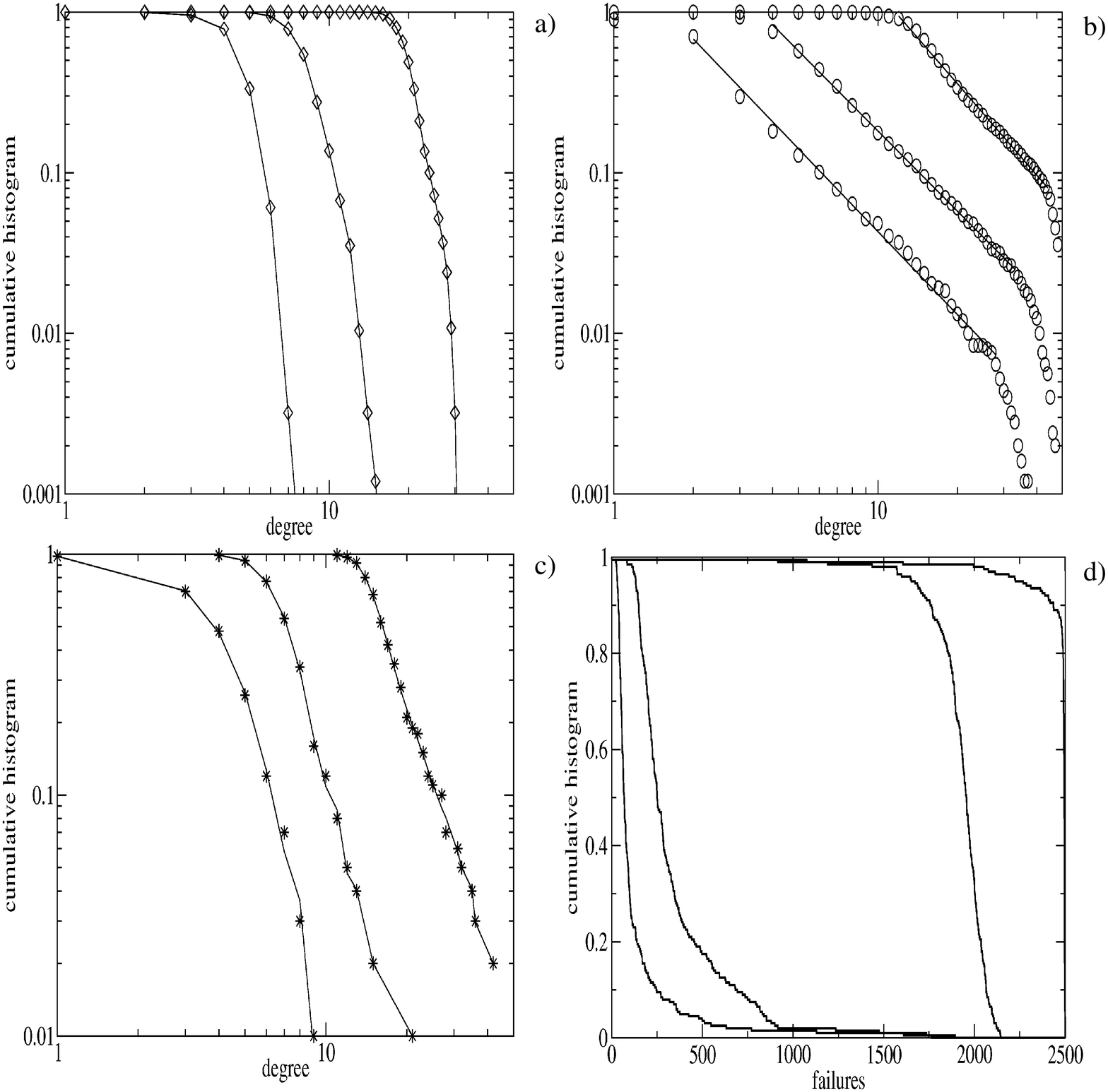}
\caption{\label{f:degree} Degree distributions for three values of $f_r= 0.3$,
$0.4$, $0.7$. Other parameter values in (a-c) as in Fig.~\ref{Topology}. d)
Cumulative failure histograms varying $f_r$ across the transition point.
\label{Fig6}}
\end{figure}

\section{Dynamics of failures} The transportation network evolves as a
consequence of the overload. As more links are added, sets of nodes will become
connected forming clusters sharing their energy. We introduce the time average
of the number of failures per node as an order parameter of the system. The
order parameter shows a transition that depends on the available energy in the
system, $f_r$. For low values of $f_r$, $\varepsilon < \lambda$, the system
evolves towards a fully connected network in which the total number of failures
is of the order of the system size (Fig.~\ref{transicion}a). The total supply in
the system is not able to sustain the needs of the nodes. If $\varepsilon >
\lambda$, a statistically failure free network is formed, accounting for a
cooperative behavior in which the network is able to effectively redistribute
the energy throughout the system. Therefore the condition $\varepsilon=\lambda$
represents a transition point.  We remark that, since the transitions occurs at
a value of  $\varepsilon$ lower than  $\varepsilon_{\rm isolated}$ (see
fig.\ref{transicion}a, inset), in our model a dynamic sharing of the resources
in a heterogeneous environment is energetically favorable. However, isolated break--down
events (a huge amount of elements fails at once, see Fig.~\ref{transicion}b)
are still present in the stable regime. These events represent global cascade
failures induced by local fluctuation, which are amplified by propagation
through the network. Global cascades are not suppressed, as it is not possible
to satisfy the total energy needs. However, as one furtherly enters the stable
regime, global cascades are less likely to occur, representing rare events that
do not prevent the average number of failure to approximate to zero.

We find interesting to analyze the transition depending on $m_r$. If $m_r$
and $f_r$ are independent, i.e., the load and the initially available energy
at each element are independent, the transition $\varepsilon=\lambda$ is
simply given by $f_r=m_r+\frac{1}{2}\sqrt{m_r}\Phi(s_{\rm max})/\Theta(s_{\rm max})$. However, we should
expect that the available energy and the load are related: elements with a
high load will also have large amount of available energy. In this case
$m_r$ depends on $f_r$ accounting for the fact that the energy availability
in each node is meant to supply the local need. If, for simplicity, we
assume that they are proportional $m_r=\mu f_r$, with $\mu\in[0,1]$ the
transition is given by $\varepsilon=\left[{\frac{\sqrt{\mu}}{2(1-\mu)}\Phi(s_{\rm max})/\Theta(s_{\rm max})}\right]^2$. In the remainder we will
assume that $\mu =1/2$.

\section{The dynamic network} A key dynamic ingredient is given by the dynamics
of the network.  The evolution of the dynamic network is driven by the failures
of the nodes and the attachment function given by Eq.~(\ref{preferential}). In
the model, failures are induced by local fluctuations in the load, afterward
propagated by the transportation network. During the transient regime, several
failures occur simultaneously leading to the initial development of the network.
After the transient, the network settles to a configuration where the average
number of failures drops to zero.  We characterize the emerging topologies in
the stationary configurations. Two factors determine the network topology: the
wiring strategy and the avaliable energy $\varepsilon$. A first characterization
of the network topology is given by the degree distribution and the average
shortest path length. By varying the parameters $\beta$ and $\gamma$ in the
wiring strategy given by Eq.~(\ref{preferential}) we obtain different classes of
networks. In Fig.~\ref{Topology} we show the degree distribution for three well
known topologies exhibited by our system for a fixed value of $\varepsilon$. The
fitted curves individuate a Gaussian degree distribution when the selection of
the target element is merely based on its relative Euclidean distance to the
failing element ($\beta=0$ and $\gamma=1$); a power-law link distribution when
the selection of the target element $j$ is based on its size $\alpha_j$
($\beta=1$ and $\gamma=0$); and finally an exponential degree distribution when
the selection of the target element $j$  is based both on its size $\alpha_j$
and relative distance to the failing element ($\beta=1$ and $\gamma=3$). To
analyze the parameter range in which different network topologies are observed,
we have measured the normalized standard deviation $\sigma/\langle k \rangle$ of
the degree distribution for different parameter values $\gamma\in[0,4]$, for
$\beta=1$. An exponential degree distribution is characterized by
$\sigma/\langle k \rangle=1$, while it increases for a power law degree
distribution. Figure \ref{sigma} shows that exponential degree distributions are
obtained for $\gamma > 2$.

\begin{figure}[tb]
\centerline{\twoimages[angle=-90,width=0.45\columnwidth]{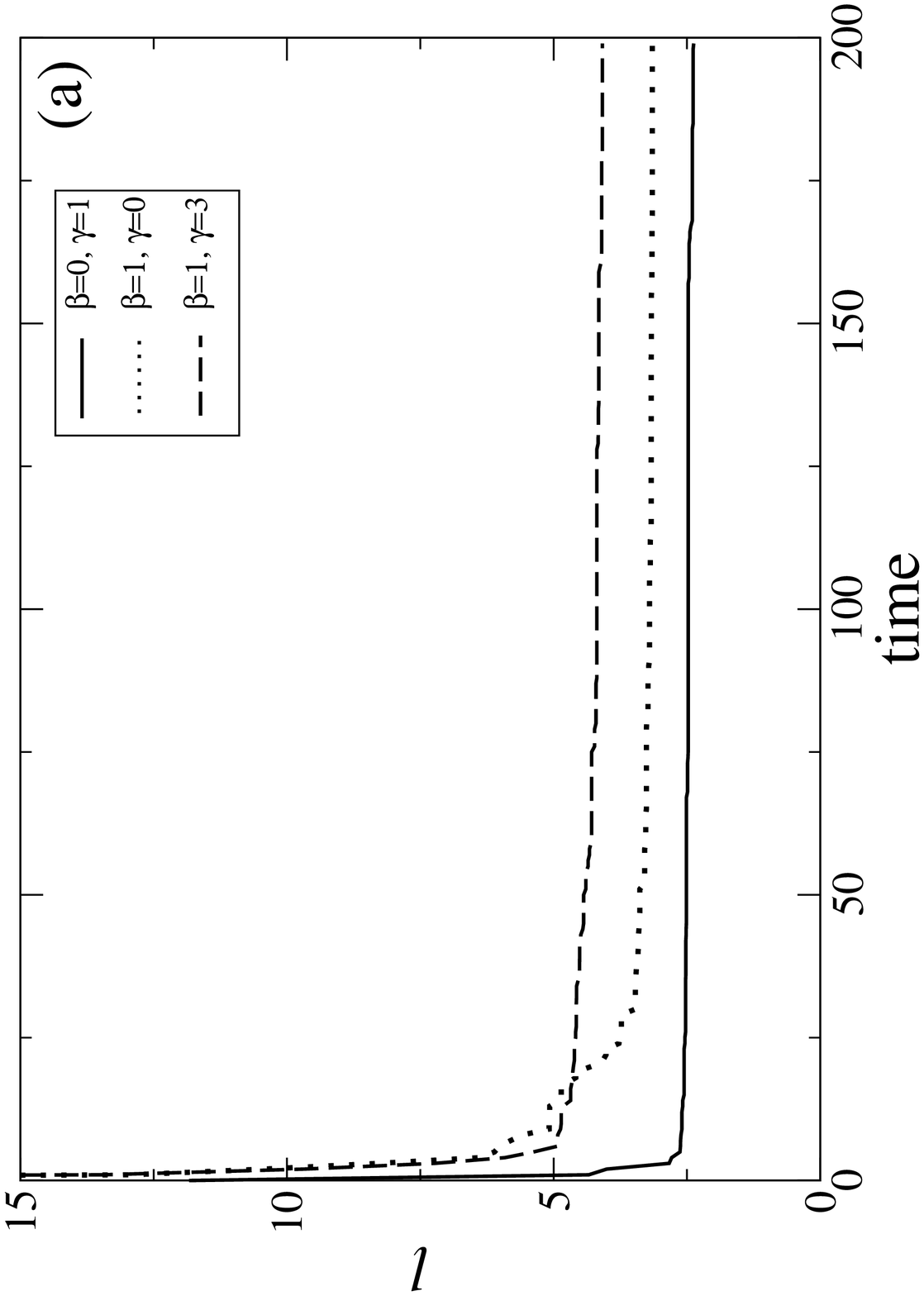}{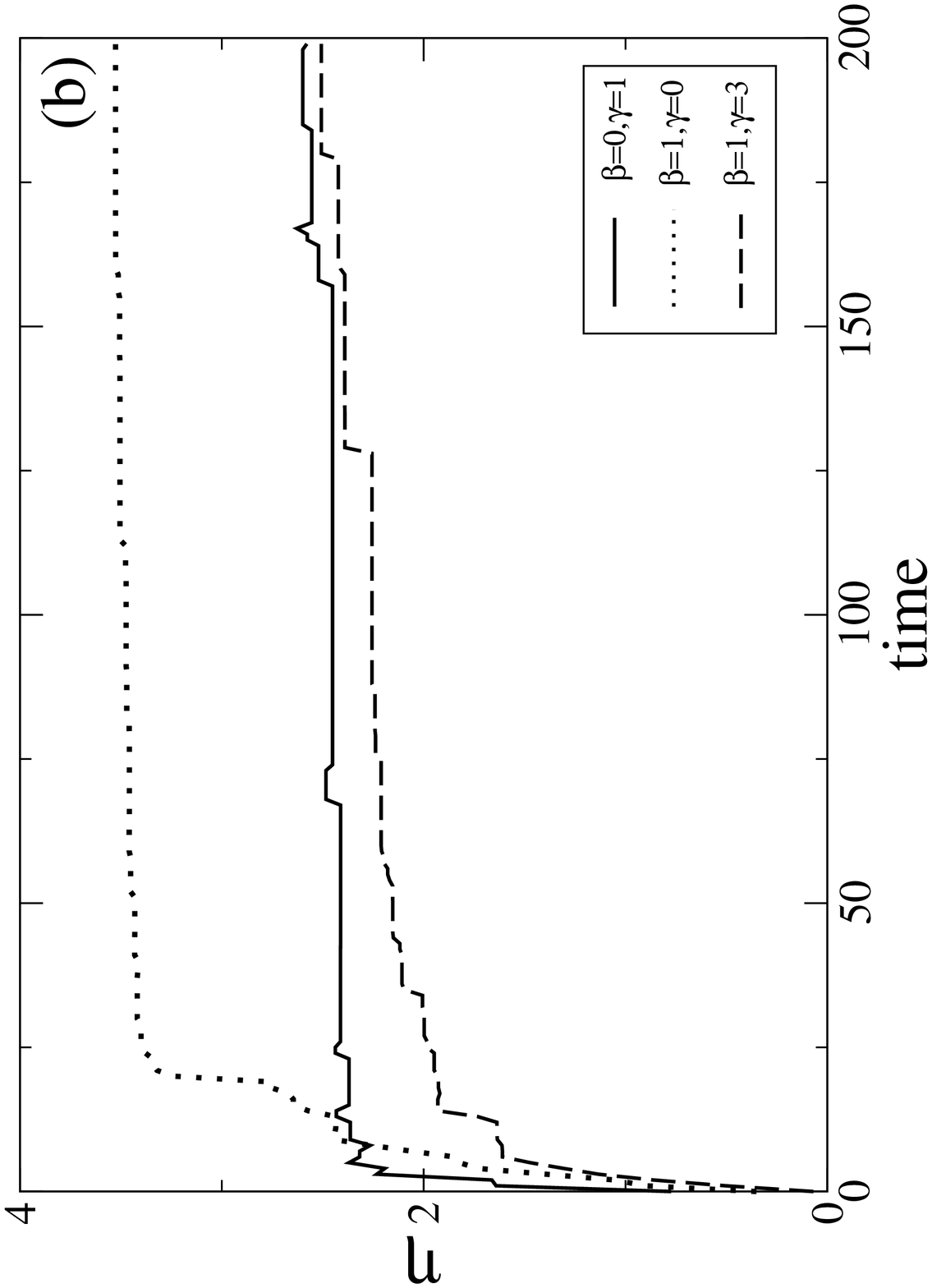}}
\caption{(a) Average shortest path length.
(b) Network efficiency. Parameter
values as in Fig.~\ref{Topology}.\label{Fig7}}
\end{figure}

Increasing the energy above its transition value ($\varepsilon=\lambda$),
the degree distribution remains qualitatively unchanged. However, a higher
number of links is required to stabilize the network as $\varepsilon$
approaches the transition value (Fig.~\ref{f:degree}). This increment in the
total number of links is due to the higher number of failures that occur
during the formation of the network closer to the transition, as is shown by
the cumulative failure histograms in Fig.~\ref{Fig6}d.

A further topological characterization is given by the average shortest path
length $l$ \cite{Watts98}
\begin{equation}
l=\frac{1}{N(N-1)} \sum_{i,j} \delta_{ij}
\label{path}
\end{equation}
where $\delta_{ij}$ is the shortest path between the elements $i$ and $j$.

During the dynamic growth of the network, $l$ decreases from infinity to a
stable finite value. For a fixed $E$ value, the minimum $l$ is obtained for
the network characterized by a power-law degree distribution
(Fig.~\ref{Fig7}a). However, this does not mean that the network possessing
a power-law degree distribution is the most efficient transporting the
energy across the system. Indeed, a possible measure of the efficiency in
the energy transport is the following:
\begin{equation}
\eta=\frac{1}{N(N-1)} \sum_{i,j} \frac{s_j}{\delta_{ij}}
\label{eff}
\end{equation}
Since in our system the largest elements act as energy reservoirs, the
definition of the efficiency $\eta$ states that the shorter is the path to
the larger elements, the more efficient is the network. Fig.~\ref{Fig7}b
shows that the network possessing an exponential degree distribution is the
most efficient in our model.

\section{Discussion} In summary we have introduced a model for the dynamic self-organization of a
transportation network, in terms of its growth and evolution. Our model
contains several features which are peculiar of the electric grid. The electric
grid is a dynamic network that evolves based on local needs, failures of local
elements, and energy sharing criteria. If the total energy is below the average
consumption, the system reaches a fully connected network; whereas if the
system has enough energy, the emerging network reflects the cooperative
behavior in which the resources are globally shared. When the cooperative
behavior is achieved, the network stops growing and reaches a configuration
with a non--trivial topology shaped by the wiring strategy and the available
total energy. Nevertheless, local fluctuations are still present once the
network is formed and keep affecting its functionality. Our model intrinsically
presents self-induced break-down events, which can be thought as representative
of real black-outs. Another remarkable feature of our model is the
heterogeneity in the initial energy distribution, modeling the geographic
inhomogeneity in the energy availability. In a heterogeneous environment, the
dynamic sharing of the resources is energetically favorable. Heterogeneity also
induces the selection of a preferred topology  as the most efficient in the
energy transport, because it displays the shortest path to the bigger elements.
The most efficient topology is characterized by an exponential degree
distribution, in agreement with the topology of the real electric grid \cite{Albert04}.

\acknowledgments We acknowledge financial support from MEC (Spain) through
project CONOCE2 (FIS2004-00953) and FIS2004-05073-C04-03.



\begin{thebibliography}{0}

\bibitem{california} http://www.spectrum.ieee.org/WEBONLY/special/aug03/blackout.html

\bibitem{IEEE}
  \Name{Kundyr P. et al.},
  \REVIEW{IEEE Transactions on Power Systems}{19}{2004}{1387}.

\bibitem{carreras}
  \Name{Carreras B.A., Lynch V. E., Dobson I. \and Newman D. E.}
  \REVIEW{Chaos}{12}{2002}{985}.

\bibitem{carreras04}
  \Name{Carreras B. A., Lynch V. E., Dobson I. \and Newman D. E.}
  \REVIEW{Chaos}{14}{2004}{643}.

\bibitem{Strogatz01}
  \Name{Strogatz S.H.}
  \REVIEW{Nature}{410}{2001}{268}.

\bibitem{Albert02}
   \Name{Albert R. \and Barab\'asi A.-L.}
   \REVIEW{Rev. Mod. Phys.}{74}{2002}{47}.

\bibitem{Dorogovtsev02}
   \Name{Dorogovtsev S. N. \and Mendes J. F. F.}
   \REVIEW{Adv. Phys.}{51}{2002}{1079}.

\bibitem{Newman03}
   \Name{Newman M.E.J.}
   \REVIEW{SIAM Review}{45}{2003}{167}.

\bibitem{Crucitti04}
   \Name{Crucitti P. \and Latora V. \and Marchiri M.}
   \REVIEW{Phys. Rev. E}{69}{2004}{045104}.

\bibitem{PastorSatorras01}
    \Name{Pastor-Satorras R. \and Vespignani A.}
    \REVIEW{Phys. Rev. Lett.}{86}{2001}{3200}.

\bibitem{Eguiluz02}
     \Name{Egu\'{\i}luz V.M. \and Klemm K.}
     \REVIEW{Phys. Rev. Lett.}{89}{2002}{108701}.

\bibitem{Eguiluz00}
   \Name{Egu\'{\i}luz V.M. \and Zimmermann M.G.}
   \REVIEW{Phys. Rev. Lett.}{85}{2000}{5659}.

\bibitem{Ito02}
   \Name{Ito J. \and Kaneko K.}
   \REVIEW{Phys. Rev. Lett.}{88}{2002}{028701}.

\bibitem{Marsili03}
   \Name{Marsili M., Vega-Redondo F. \and F. Slanina}
   \REVIEW{Proc. Natl. Acad. Sci. U.S.A.}{101}{2003}{1439}.

\bibitem{Anghel04}
   \Name{Anghel M., Toroczkai Z. Bassler, K.E. \and Korniss G.}
   \REVIEW{Phys. Rev. Lett.}{92}{2004}{058701}.

\bibitem{bianconi04}
   \Name{Bianconi G. \and Marsili M.}
   \REVIEW{Phys. Rev. E}{70}{2004}{035105}.

\bibitem{Zimmermann04}
   \Name{Zimmermann M.G., Egu\'{\i}luz V.M. \and San Miguel M.}
   \REVIEW{Phys. Rev. E}{69}{2004}{065102(R)}.

\bibitem{zipf}
   \Name{Zipf G.K}
   \Book{Human behavior and the principle of least effort}
   \Publ{A-W, Cambridge MA}
   \Year{1949}.

\bibitem{ruido} We chose alternatively deterministic (logistic map: \Name{May R.M.} \REVIEW{Nature}{261}{1976}{456}) or
stochastic (white) fluctuation to emphasize possible deterministic effects of
the local dynamics on the network behavior. The nature of the fluctuations
resulted not to qualitatively affect the reported results, which relay on the
use of the logistic map.

\bibitem{Yook02}
   \Name{Yook S.-H., Jeong H. \and Barab\'asi A.-L.}
   \REVIEW{Proc. Natl. Acad. Sci. USA}{99}{2002}{13382}.

\bibitem{Watts98}
   \Name{Watts D.J. \and Strogatz S.H.}
   \REVIEW{Nature}{393}{1998}{409}.

 \bibitem{Albert04}
   \Name{Albert R., Albert I. \and Nakarado G.L.}
   \REVIEW{Phys. Rev. E}{69}{2004}{025103}.

\end{thebibliography}
\end{document}